%% file: index.tex
\documentclass[12pt]{article}

\usepackage{graphicx}
\usepackage{amssymb}

\newcommand{\be}{\begin{equation}}
\newcommand{\ee}{\end{equation}}
\newcommand{\bea}{\begin{eqnarray}}
\newcommand{\eea}{\end{eqnarray}}
\newcommand{\nn}{\nonumber}
\newcommand{\Tr}{{\rm Tr}}
\newcommand{\diag}{{\rm diag}}
\newcommand{\sgn}{{\rm sgn}}
\newcommand{\R}{{\mathbb R}}
\newcommand{\CP}{\mathbb{CP}}
\newcommand{\WCP}{\mathbb{WCP}}
\newcommand{\U}{{\bf U}}
\newcommand{\Lcal}{{\mathcal L}}
\newcommand{\M}{{\mathcal M}}
\newcommand{\W}{{\mathcal W}}
\newcommand{\Ocal}{{\mathcal O}}
\newcommand{\rot}[3]{\left[{#1}\atop{#2}\right]_{#3}}
\newcommand{\binom}[2]{\left({#1}\atop{#2}\right)}

\makeatletter
  
  \@addtoreset{equation}{section}
\makeatother

\begin{document}

%
%
\begin{titlepage}

\setcounter{page}{0}
\renewcommand{\thefootnote}{\fnsymbol{footnote}}

\begin{flushright}
YITP-99-50 \\
hep-th/9908120
\end{flushright}

\vspace{15mm}
\begin{center}
{\large\bf
Supersymmetric Index and
 $s$-rule for Type IIB Branes
}

\vspace{15mm}
{\large
Kazutoshi Ohta\footnote{e-mail address:
kohta@yukawa.kyoto-u.ac.jp}
}\\
\vspace{10mm}
{\em Yukawa Institute for Theoretical Physics, Kyoto University,
Kyoto 606-8502, Japan} \\
\end{center}
\vspace{15mm}
\centerline{{\bf{Abstract}}}
\vspace{5mm}
We investigate the supersymmetric index of $N{=}2,3$ $SU(n)$ supersymmetric
Yang-Mills Chern Simons theories at level $k$ by using the brane
configuration with a $(p,q)$5-brane.  We can explain that the
supersymmetry breaking occurs when $k<n$ in terms of the $s$-rule for Type
IIB branes.  The supersymmetric index coincides with the number of the
possible supersymmetric brane configurations. We also discuss a
construction of a family of theories which have the same supersymmetric
index.
\end{titlepage}
\newpage

\renewcommand{\thefootnote}{\arabic{footnote}}
\setcounter{footnote}{0}

%
%
\section{Introduction}

A supersymmetric index (Witten index) $\Tr(-1)^F$ is a useful
criterion of dynamical supersymmetry breaking in supersymmetric field
theories \cite{Witten2}. The index is given by a difference of the
number of bosonic and fermionic zero energy states and it counts the
number of the supersymmetric vacua. When $\Tr (-1)^F \neq 0$
supersymmetry is not spontaneously broken since if pairs of bosonic
and fermionic zero energy states gains a non-zero energy at least
$\Tr(-1)^F$ states remain as zero energy states. Conversely, however,
for the case of $\Tr (-1)^F = 0$ we can not conclude whether
supersymmetry is broken or not. But it indicates a possibility of
spontaneous supersymmetry breaking.

The indexes for various supersymmetric theories have been
computed. Recently, the index for supersymmetric gauge theory with a
mass gap, which is $N{=}1$ supersymmetric Yang-Mills Chern-Simons
theory with gauge group $G$, is given by Witten \cite{Witten}. His
result shows that $\Tr (-1)^F
\neq 0$ if $|k|\geq h/2$, where $h$ is the dual Coxeter number of $G$,
and suggests that dynamical symmetry breaking occurs for $|k|<h/2$.

Three-dimensional supersymmetric Yang-Mills Chern-Simons theory can be
realized on the brane configuration with a $(p,q)$5-brane in Type IIB
superstring theory \cite{KOO,Ohta,LLOY,KO}. For $N{=}1$ supersymmetric theory
without any extra matters, the corresponding brane configuration has
not been well understood. But $N{=}2,3$ supersymmetric theories are easy
to handle by the branes. The moduli space of vacua of $N{=}2,3$ theory
can be explained by the brane configuration \cite{Ohta}. So we focus
only $N{=}2,3$ theory throughout the present paper.

In this paper, we extend the Witten's computation of the index to
$N{=}2,3$ $SU(n)$ super Yang-Mills Chern-Simons theory at level $k$
and show that this index is well interpreted by using the Type IIB
brane configuration. Especially, the dynamical supersymmetry breaking
in some region of $k$ is explained by the so-called ``$s$-rule''
(supersymmetric rule) for the branes. Moreover, when the index is
non-zero we show that the index exactly coincides of the number of
possible supersymmetric brane configurations.

The organization of this paper is as follows. In section 2, we briefly
review a microscopic computation of the index by Witten. The
computation can straightforwardly extend to $N{=}2,3$ extend
supersymmetry. We will give the formulae of the indexes. In section 3,
we explain about the $s$-rule for the branes and generalize it to the
stretched D3-branes between two different types of $(p,q)$5-branes. We
next show in section 4 that if this generalized $s$-rule is applied to
the $N{=}2,3$ Yang-Mills Chern-Simons brane configuration, the
configuration for $k<n$ can not keep the $s$-rule, that is, the
configuration is not supersymmetric. These results are consistent with
the computation of the index. We will give the exact formula of the
index by counting the number of possible supersymmetric configuration
from M-theoretical point of view. We also discuss the relation between
a supersymmetric quantum mechanics and the computation of the index in
terms of the branes. In the subsequent subsection, we construct a
family of theories which have the same index by using some brane
dynamics. Finally, section 5 is devoted to a summary of our results
and discussions of further problems.

%
%
\section{Brief review of the microscopic computation of the index}
\label{index comp}

We first consider $N{=}1$ $SU(n)$ supersymmetric Yang-Mills
Chern-Simons theory on $\R\times T^2$. The computation is done by a
Born-Oppenheimer approximation. In this approximation we take a small
volume limit of the torus. If $g$ and $r$ denote the gauge
coupling and the radius of the torus, the mass of the vector
multiplet $kg^2$ is much smaller than the Kaluza-Klein mass of order
$1/r$. We will consider the quantizing of the ``zero energy'' states,
whose energy are of order $kg^2$ by means of the approximation.

Now let us consider the moduli space of a flat $G$-connection on
$T^2$, which is a zero energy classical gauge field configuration. The
moduli space $\M$ of flat $G$-connections on $T^2$ is given by
\be
\M=(\U\times\U)/W,
\ee
where $\U$ is the maximal torus of $G$ and $W$ is the Weyl group. For
$G=SU(n)$, $\M$ becomes simply the complex projective space $\CP^{n-1}$
\cite{FMW}.

We next consider ``zero modes'' for the gluino fields of positive and
negative chirality $\lambda_{+}$ and $\lambda_{+}$, which have at most
energy of order $kg^2$. We assume the zero modes of $\lambda_\pm$ take
of the form:
\be
\lambda_\pm=\sum_{a=1}^{r}\eta_\pm^a T^a,
\ee
where the $T^a, a=1,\ldots ,r$ are a basis of the Lie algebra of
$\U$ and $\eta_\pm^a$ are fermionic constants, which obey the
following canonical anti-commutation relations
\be
\{\eta_+^a,\eta_-^b\}=\delta^{ab}, \quad \{\eta_+,\eta_+\}=\{\eta_-,\eta_-\}=0.
\label{anticomm}
\ee
The $\eta_+$ and $\eta_-$ can be regarded as creation and annihilation
operators, so we now introduce ground states annihilated by the
$\eta_\pm^a$
\be
\eta_\pm|\Omega_\pm\rangle=0.
\ee
These two states are related each other by
\be
|\Omega_{+}\rangle=\prod_{a=1}^{r}\eta^a_{+}|\Omega_{-}\rangle.
\ee

The anti-commutation relations (\ref{anticomm}) can be regarded as the
Clifford algebra on $\M$. Therefore, the Hilbert space made by
quantizing the fermion zero modes maps to the space of spinor fields
on $\M$. Since $\M$ is a complex manifold, the spinor field on $\M$ is
simply represented by the form with values in $K^{1/2}$, where $K$ is
the canonical line bundle of $\M$. So if we consider a general state
in the fermionic space
\be
\eta_-^{a_1}\cdots\eta_-^{a_q}|\Omega_+\rangle,
\ee
this is regarded as the $(0,q)$-form on $\M$ with values in $K^{1/2}$. 
More precisely speaking, the forms on $\M$ take values not only in
$K^{1/2}$ but also in a line bundle $\W$, since $|\Omega_+\rangle$ is
regarded as a $(0,0)$-form on $\M$ with values in $\W$.

To determine the line bundle $\W$ let us consider the canonical
quantization of the Yang-Mills Chern-Simons theory. The momentum
conjugate to $A_i^a$ is given by
\be
\Pi_i^a=\frac{1}{g^2}F_{0i}^a-\frac{k}{4\pi}\epsilon_{ij}A_j^a.
\ee
Writing formally $\Pi_i^a=-i\delta/\delta A_i^a$, we have
\be
\frac{1}{g^2}F_{0i}^a=-i\frac{D}{DA_i^a},
\ee
where
\be
\frac{D}{DA_i^a} = \frac{\delta}{\delta A_i^a}
+i\frac{k}{4\pi}\epsilon_{ij}A_j^a
\label{connection}
\ee
is a connection on the line bundle $\W$. The connection form
$\frac{k}{4\pi}\epsilon_{ij}A_j^a$ means that the line bundle over
$\M$ is $\Lcal^k$, where $\Lcal$ is the basic line bundle. The
supercharges of the theory are written in terms of the connection
\bea
Q_+ & = & \frac{1}{g^2}\int_{T^2}\Tr F_{0z}\lambda_{-}
 = \int_{T^2}\Tr\lambda_{-} \frac{D}{DA_{\bar{z}}},\\
\label{supercharge1}
Q_- & = & \frac{1}{g^2}\int_{T^2}\Tr F_{0\bar{z}}\lambda_+
 = \int_{T^2}\Tr\lambda_{+} \frac{D}{DA_z}.
\label{supercharge2}
\eea
Therefore, the supercharges $Q_+$ and $Q_-$ can be identified with the
$\bar{\partial}$ and $\bar{\partial}^\dagger$ operators, which is a
decomposition of a Dirac operator acting on spinors valued in
$\W=\Lcal^k$.

Since these two operators obey
\be
\{\bar{\partial},\bar{\partial}^\dagger\}=H,\quad
\bar{\partial}^2=(\bar{\partial}^\dagger)^2=0,
\ee
where $H$ is the Hamiltonian, the space of supersymmetric ground
states is given by the cohomology
\be
\bigoplus_{i=0}^{n-1} H^i\left(\M,\W\otimes K^{1/2}\right).
\ee
For $G=SU(n)$, $\M\simeq\CP^{n-1}$. The basic line bundle over $\M$ is
$\Lcal=\Ocal(1)$ and the canonical bundle of $\M$ is
$K=\Lcal^{-n}$. Therefore, the above cohomology groups are rewritten as
\be
\bigoplus_{i=0}^{n-1} H^i\left(\CP^{n-1},\Lcal^{k{-}n/2}\right),
\ee
where we use $\W=\Lcal^k$.
Thus, the supersymmetric index is
\be
I_{N{=}1}(k)=\sum_{i=0}^{n-1}(-1)^i \dim H^i\left(\CP^{n-1},\Lcal^{k{-}n/2}\right).
\ee
These cohomology groups are computed by Serre \cite{Serre} and their
dimensions are as follows
\be
\dim H^i\left(\CP^{n-1},\Lcal^r\right)=\left\{
\begin{array}{ll}
0 & \mbox{for $0{<}i{<}n{-}1$}\\
0 & \mbox{for $i{=}n{-}1$, $r{>}{-}n$}\\
0 & \mbox{for $i{=}0$, $r{<}0$}\\
\binom{n{+}r{-}1}{r} & \mbox{for $i{=}0$, $r{\geq}0$}
\end{array}
\right.,
\ee
where $\binom{n{+}r{-}1}{r}$ is a binomial coefficient and means the
dimension of the vector space of degree $r$ homogeneous polynomials in
the $n$ homogeneous coordinates of $\CP^{n-1}$.  Using this formula we
obtain the supersymmetric index of the $N{=}1$ Yang-Mills Chern-Simons
theory,
\be
I_{N{=}1}(k)=\left\{
\begin{array}{ll}
0 & \mbox{for $|k|{<}n/2$}\\
\binom{k{+}n/2{-}1}{k{-}n/2} & \mbox{for $|k|{\geq}n/2$}
\end{array}
\right.,
\ee
where for the case of $k\leq -n/2$ we use Serre duality
\be
H^{n-1}(\CP^{n-1},\Lcal^r)\simeq H^0(\CP^{n-1},\Lcal^{-n-r}) \quad \mbox{for $r\leq -n$}.
\ee

We now extend this computation of the index to the case of $N{=}2,3$
supersymmetric Yang-Mills Chern-Simons theory in
three-dimensions. First, we note that the $N{=}2$ vector multiplet
consists of one spin 1 vector boson $A_\mu$, two spin 1/2 spinors
$\lambda_1,\lambda_2$ and one spin 0 real adjoint scalar $X$. Since
the adjoint scalar has the mass $kg^2$, the Coulomb branch of this
theory is completely lifted. Therefore, the presence of the adjoint
scalar does not affect the index. So, the computation of the index is
modified by two gluino fields as
\bea
I_{N{=}2}(k) &=& \sum_{i=0}^{n-1}(-1)^i \dim H^i\left(\CP^{n-1},\W\otimes K^{1/2}\otimes K^{1/2}\right)\nn\\
& = & \sum_{i=0}^{n-1}(-1)^i \dim H^i\left(\CP^{n-1},\Lcal^{k{-}n}\right).
\eea
Similarly, for the $N{=}3$ theory, the massive vector multiplet
contains one spin 1 vector boson $A_\mu$, three spin 1/2 spinors
$\lambda_1,\lambda_2,\lambda_3$, three spin 0 real adjoint scalars
$X_1,X_2,X_3$ and one spin -1/2 spinor $\chi$. In this case, there is
also no Coulomb branch moduli. The gluino fields contain one opposite
spin field, so the contribution of one of spin 1/2 fields is canceled by
the spin -1/2 field $\chi$.
\bea
I_{N{=}3}(k) & = & \sum_{i=0}^{n-1}(-1)^i 
\dim H^i\left(\CP^{n-1},\W\otimes K^{3/2}\otimes K^{-1/2}\right)\nn\\
&=& \sum_{i=0}^{n-1}(-1)^i \dim H^i\left(\CP^{n-1},\Lcal^{k{-}n}\right)
\eea
Therefore, the index of $N{=}3$ theory coincides with the $N{=}2$ one. 
We can again obtain the index of the $N{=}2,3$ theories by the Serre's
formula.
\be
I_{N{=}2,3}(k)=\left\{
\begin{array}{ll}
0 & \mbox{for $0{<}k{<}n$}\\
\binom{k{-}1}{k{-}n}
=\binom{k{-}1}{n{-}1}
& \mbox{for $k{\geq}n$}
\end{array}
\right..
\label{index}
\ee
For $k\leq0$ we also compute the index by using the Serre duality,
which is non-zero valued for any $n$, but the field theoretical
meaning of this duality has not been understood at the present. So, we
assume that $k$ is a positive integer in the following.  We will
discuss the relation between the index and the brane configuration in
the following section.

\section{The $s$-rule for Type IIB branes}
\label{s-rule}

In this section, we explain about the $s$-rule for branes in string
theory. The $s$-rule is a phenomenological rule of brane dynamics,
which is first proposed by Hanany and Witten \cite{HW}, and is needed
in order to make supersymmetric configurations. For example, a
NS5-brane and a D5-brane, which are completely twisted in the
configuration space, can be supersymmetrically connected by only one
D3-brane. If we use this rule we can find the exact correspondence
between the brane configuration and the supersymmetric vacua of field
theories. Some explanations on this rule are given from various point
of view in Refs.~\cite{OV, HOO, NOYY, BGS, BG}.  If we map the
D3-brane between the NS5-brane and the D5-brane to an M2-brane between
two M5-branes by U-duality, we obtain the following rule in M-theory:

{\it A configuration in which two completely twisted M5-brane are
connected by more than one M2-brane is not supersymmetric.}

We can obtain all of other supersymmetric brane configurations keeping
the $s-$rule in various string theory from this M-theoretical rule by
string duality. So we now consider the generalization of the $s$-rule
for the D3-branes between two different type of the
$(p,q)$5-brane. The $(p,q)$5-brane in Type IIB theory is described by
a single M5-brane which wrapping simultaneously on two cycles of the
compactified torus of M-theory. The wrapping number of the each cycles
corresponds to charges of the $(p,q)$5-brane. This wrapping cycle is
denoted by $p\alpha +q\beta$, where $\alpha$ and $\beta$ stand for the
two independent cycles of the torus. If we introduce another
$(p',q')$5-brane the corresponding M5'-brane similarly wraps on a
$p'\alpha +q'\beta$ cycle. The M5-brane and the M5'-brane meet
$|pq'-qp'|$ times on the torus, where $|pq'-qp'|$ is an intersection
number of cycles $p\alpha +q\beta$ and $p'\alpha +q'\beta$. Since only
one M2-brane can be attached on each intersecting point if we use the
above $s$-rule in M-theory, the maximal number of the M2-branes
between M5- and M5'-brane must be $|pq'-qp'|$ to preserve
supersymmetry. If the number of M2-branes is more than $|pq'-qp'|$, we
can not arrange the M2-branes in a supersymmetric way.

By using string duality, the M5- and M5'-brane maps to the $(p,q)$5-
and $(p',q')$5-brane in Type IIB theory. So we find that if the number
of the D3-branes between the $(p,q)$5- and $(p',q')$5-brane is more
than $|pq'-qp'|$ then the configuration is not
supersymmetric.\footnote{If $pq'-qp'$ is negative, it means that the
stretched D3-branes are anti-D3-branes which have opposite orientation 
and charge.}

%
%
\section{Comparison with supersymmetric Yang-Mills Chern-Simons theory configuration}

In this section, we apply the result of the previous section to the
brane configuration of supersymmetric Yang-Mills Chern-Simons theory
and compare with the computation of the index. $N{=}2,3$
supersymmetric $SU(n)$ Yang-Mills Chern-Simons theories at level $k$ are
realized on the generalized Hanany-Witten type configurations, in
which $n$ D3-branes are suspended between an NS5-brane and a
$(p,q)$5-brane \cite{KOO}. The coefficient of the Chern-Simons term,
that is, the level of Chern-Simons theory is given by $k=p/q$ in this
brane setup. For non-Abelian gauge theory this coefficient should be
an integer due to the quantization condition. So, we set $p=k$ and
$q=1$ in the pages that follow to avoid subtlety.

Moreover, in order to compare with the result of the index
computation, we only consider theories without any extra matter except for
vector multiplet. For $N{=}2$ Yang-Mills Chern-Simons theory, the
configuration is\footnote{For derivation and notation, see
Ref.~\cite{KOO}.}
\be
\begin{array}{l}
\mbox{NS5}(012345)\\
\mbox{D3}(012|6|)\\
\mbox{$(k,1)$5}\left(01278\rot{5}{9}{-\theta}\right)
\end{array},
\label{N=2}
\ee
and for $N{=}3$,
\be
\begin{array}{l}
\mbox{NS5}(012345)\\
\mbox{D3}(012|6|)\\
\mbox{$(k,1)$5}
\left(012\rot{3}{7}{\theta}\rot{4}{8}{\theta}\rot{5}{9}{-\theta}\right)
\end{array}.
\label{N=3}
\ee

\subsection{Supersymmetric index from branes}

We first apply the generalized $s$-rule to the Type IIB brane
configuration which describes $SU(n)$ Yang-Mills Chern-Simons theory
at level $k$. In this configuration one of the 5-brane is an
NS5-brane, namely, a $(0,1)$5-brane. Another is a
$(k,1)$5-brane. These 5-branes are completely twisted in the $N{=}2$
and $N{=}3$ configuration (\ref{N=2}) and (\ref{N=3}). So, the
supersymmetric configuration is restricted by the $s$-rule. The
intersection number of these 5-branes in M-theory is $k$. Therefore,
for supersymmetry, the maximal number of the D3-branes must be $k$.

Therefore, we find that the configuration is supersymmetric if $n\leq
k$ and the situation for $n>k$ violates the $s$-rule and spontaneously
breaks supersymmetry. This exactly agree with the computation of the
index, that is, $I_{N{=}2,3}(k)\neq 0$ for $n\leq k$ and
$I_{N{=}2,3}(k)=0$ for $n>k$.

We next consider how the value of the index itself is described in the
brane configuration. The value of the index means
the number of the possible supersymmetric vacua. So, we can expect
that the non-zero value of the index coincides with the number of
possible supersymmetric configurations of branes.

In order to count the number of supersymmetric configuration, we again
lift the Type IIB configuration to M-theory. The NS5-brane is an
M5-brane wrapping on a cycle $\beta$ and the $(k,1)$5-brane is an
M5-brane wrapping on a cycle $k\alpha +\beta$ in M-theory. These
M5-branes intersect $k$ times on the torus. Therefore the M2-brane can
be attached on $k$ intersecting points, but can not be attached on the
same point by the $s$-rule. If $n> k$, the $n$ M2-branes can not be
arranged without violating the $s$-rule. So, supersymmetry is
broken. For $n\leq k$ the number of possible supersymmetric
configurations coincides with the number of choices of $n$ points within
$k$ points, that is, $\binom{k}{n}$.

This number seems not to be the same with the index
(\ref{index}). However, precisely speaking, the gauge group $G$ on the
$n$ D3-branes, which we are considering, is $U(n)$ rather than
$SU(n)$. Since the computation of the index in section \ref{index
comp} is for $G=SU(n)$, this discrepancy occurs. To compare with the
index for $SU(n)$ gauge theory, we must fix one of positions of the
M2-brane, which corresponds to a phase of overall $U(1)$ factor of
$U(n)$. If we fix the position of one M2-brane, residual ways of the
supersymmetric arrangement of M2-branes is $\binom{k-1}{n-1}$. This
number exactly agrees with the index (\ref{index}).

\begin{figure}
\centerline{
\includegraphics[scale=0.8]{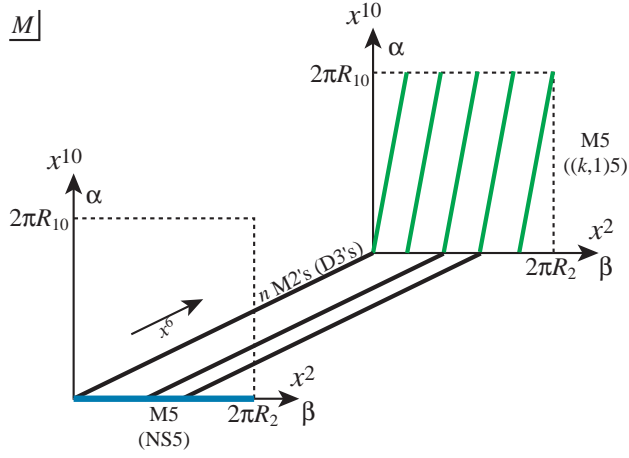}
}
\caption{The brane configuration in M-theory for $n\leq k$.
$n$ M2-branes can be attached one by one only on $k$ positions.}
\label{M config}
\end{figure}

A similar derivation of the supersymmetric index by counting possible
supersymmetric configurations is discussed in
the case of the M-theory description of $N{=}1$ supersymmetric
Yang-Mills theory in Ref.~\cite{Witten3}.

As discussed in Ref.~\cite{KOO}, the positions of M2-branes correspond
to vevs of the Wilson line operator $W_2$ along the
$x^2$-direction. In the supersymmetric configuration, all of M2-brane
positions are different. So, we find that the vevs of the Wilson line
operator must take the following form
\be
\langle W_2 \rangle 
= \diag\left(e^{2\pi i m_1/k}, e^{2\pi i m_2/k},
 \ldots , e^{2\pi i m_n/k}\right),
\ee
up to over all phase, where $m_i$ are different positive integers
which satisfy $0\leq m_1<m_2<\cdots<m_n<k$. Therefore, we can conclude
that in the supersymmetric phase, gauge symmetry of $N{=}2,3$ $U(n)$
Yang-Mills Chern-Simons theory is broken to $U(1)^n$ by the vev of the
Wilson line operator. (For $G=SU(n)$, broken to $U(1)^{n-1}$.)

\subsection{Supersymmetric quantum mechanics on the brane}

Let us next consider a dual description of the above Type IIB brane
configuration which describes supersymmetric Yang-Mills Chern-Simons
theory. That picture make easy to learn about the connection to the
computation of the index.

We first consider a Maxwell Chern-Simons theory for simplicity. The
corresponding brane configuration is that a single D3-brane is
suspended between a NS5-brane and a $(k,1)5$-brane. If we consider the
long wavelength limit of the Maxwell Chern-Simons Lagrangian, in which
we drop all spatial derivatives, we obtain
\be
L=\frac{1}{2g^2}\dot{A}_i^2+\frac{k}{2}\epsilon^{ij}\dot{A}_iA_j,
\ee
where the gauge coupling $g$ is give by the sting coupling $g_s$ and
the difference between two 5-branes $L$ as $1/g^2=L/g_s$.  This
Lagrangian has exactly the same form as the Lagrangian for a
non-relativistic charged particle with mass $1/g^2$ moving in the
plane in the presence of a external magnetic flux $k$
perpendicular to the plane. Therefore, the Maxwell Chern-Simons system
has the same canonical structure with the Landau problem.

\begin{figure}
\centerline{
\includegraphics[scale=0.7]{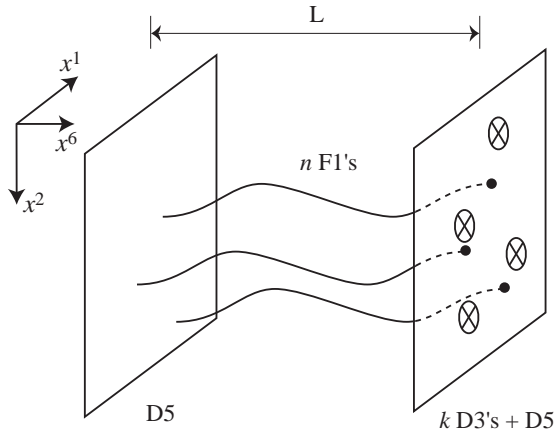}
}
\caption{The Landau system
on the D3-D5 bound state.}
\label{Landau config}
\end{figure}

On the other hand, if we take T-dual along the $x^1$- and
$x^2$-direction and S-dual in Type IIB theory to the $N{=}2$
configuration (\ref{N=2}), we have
\be
\begin{array}{l}
\mbox{D5}(012345)\\
\mbox{F1}(0|6|)\\
k\mbox{D3}\left(078\rot{5}{9}{-\theta}\right) 
\oplus \mbox{D5} \left(01278\rot{5}{9}{-\theta}\right)
\end{array},
\ee
where $k\mbox{D3}\oplus\mbox{D5}$ is a bound state of $k$ D3-branes
and a D5-brane. On the D3-D5 bound state, $k$ D3-branes are regarded
as $k$ magnetic flux on the D5-brane. Therefore, on the D3-D5 bound
state the end of the fundamental string stretched from the D5-brane
looks like a charged particle in the $k$ magnetic flux. (See
Fig. \ref{Landau config}.) Since the gauge field configuration $A_i$
maps to the position of the fundamental string $X_i$ by T-duality, we
obtain really the (supersymmetric) quantum mechanical Lagrangian of
the Landau problem for the position of the string on $(x^1,x^2)$-plane
after using T-duality and S-duality
\be
L=\frac{1}{2}\left(\frac{L}{2\pi l_s^2}\right)\dot{X}_i^2
+\frac{\widetilde{k}}{2}\epsilon^{ij}\dot{X}_iX_j,
\label{Landau}
\ee
where $l_s$ is a string length and $\widetilde{k}=\frac{k}{2\pi l_s^2
g_s}$. This is a Lagrangian for the particle with mass which is string
tension times length $L$ in magnetic flux which is proportional to $k$
as expected from the dualized brane configuration.

If we now extend the above situation to $SU(n)$ gauge theory, $n$ becomes
the number of particles. Due to the $SU(n)$ restriction the sum of the
particle positions must vanish. The moduli space of the $n$-tuple of
such the particle positions on the torus is known as a copy of complex
projective space $\CP^{n-1}$
\cite{FMW}, which is a phase space of the quantum mechanical system
(\ref{Landau}).

After all, analysis of the vacuum structure of supersymmetric
Yang-Mills Chern-Simons theory replaces the quantization of the
supersymmetric quantum mechanics on the D3-D5 bound state by using
U-duality. This is the same thing with the microscopic computation of
the index in section \ref{index comp}. The covariant derivative
(\ref{connection}) and the supercharges (\ref{supercharge1}) and
(\ref{supercharge2}) can be regarded as operators acting on the
supersymmetric quantum mechanics of the Landau problem.

Finally, we comment on a non-commutative nature on the torus. In limit
of $L\ll l_s$, the Lagrangian (\ref{Landau}) becomes
$L=\frac{\widetilde{k}}{2}\epsilon^{ij}\dot{X}_iX_j$. This is first
order in time derivatives, so the two coordinates $X_1$ and $X_2$ are
canonically conjugate to one another, that is,
\be
\left[X_i,X_j\right]=i\epsilon_{ij}/\widetilde{k}.
\ee
Thus, the coordinates of the ends of open strings are non-commutative.
We can understand this property as the non-commutative torus by
turning on a non-zero B-field background using a gauge transformation \cite{CDS,DH}. 
This non-commutative nature is closely related to the dynamics of
Chern-Simons theory.

\subsection{Duality and mirror relations}

In this subsection we present a family of theories with a Chern-Simons
term, which have the same supersymmetric index. We first consider an
exchange of 5-branes in the $x^6$-direction. When two different types
of twisted 5-branes cross in the $x^6$-coordinate and exchange
positions, some D3-branes are annihilated or created by the
Hanany-Witten transition \cite{HW}. For example, if $n$ D3-branes are
stretched between the NS5-brane and the $(k,1)$5-brane ($n<k$), then
the number of stretched D3-branes becomes $k{-}n$ after the
transition.

This phenomenon is also simply explained from the M-theoretical point
of view \cite{KOO}. In M-theory, the Hanany-Witten transition is
creation or annihilation of a single M2-brane between two twisted
M5-branes, which can not avoid each other in the configuration
space. In the case of $n<k$, M2-branes can be attached on the $n$
different positions within the $k$ allowed positions on the
M5-brane. After exchanging the positions of the two M5-branes in the
$x^6$-coordinate, $n$ M2-branes are disappeared and new M2-branes are
created on $k{-}n$ opening positions.

Thus, we find that by the Hanany-Witten transition the gauge group of
$N{=}2,3$ supersymmetric $U(n)$ Yang-Mills Chern-Simons theory at
level $k$ becomes $U(k{-}n)$. (If $G=SU(n)$, then
$\hat{G}=SU(k{-}n{+}1)$ since we must fix one of the M2-branes.) On
this transition the number of the possible supersymmetric
configuration, namely, the index, has not been changed because of the
equivalence of the combination $\binom{k}{n}=\binom{k}{k{-}n}$.

Another operation to the brane configuration is S-duality in Type IIB
theory. If we apply the S-duality to the brane configuration which
describes supersymmetric Yang-Mills Chern-Simons theory at level $k$,
we have a self-dual model at level $-1/k$ as a worldvolume effective
theory \cite{KOO}. Since this S-dual transforation can be understood
as just a coordinate flip of $x^2$ and $x^{10}$ in M-theory, there is
no difference between the configurations of the Yang-Mills
Chern-Simons theory and self-dual model themselves. So, we expect that
these theories have the same supersymmetric index.

In this way, we can construct the family of theories with the same
index by using the operations in superstring theory. For $N{=}2,3$
supersymmetric theories, $U(n)$ and $U(k{-}n)$ Yang-Mills Chern-Simons
theory at level $k$ and $U(n)$ and $U(k{-}n)$ valued self-dual models at
level $-1/k$ have all the common supersymmetric index. It suggests
that these theories are related each other by duality and mirror
symmetry of three-dimensional supersymmetric field theory.

\section{Conclusion and discussions}

We have investigated in this paper the relation between the
supersymmetric index of three-dimensional supersymmetric Yang-Mills
Chern-Simons theory and the brane configuration in Type IIB
superstring theory. We have found that when the index is non-zero
valued, the corresponding brane configuration is
supersymmetric. Conversely, when the index vanishes, the configuration
violates the $s$-rule and becomes non-supersymmetric. In addition, we
have found that the number of the possible supersymmetric
configuration exactly coincides with the value of the index. These
results are considered as an explanation of the $s$-rule for the brane
dynamics from the point of view of the worldvolume effective theory.

The computation of the index can be generalized to other gauge groups
$G$ of rank $r$ \cite{Witten}. The moduli space $\M$ of the flat
$G$-connection is a weighted projective space
$\WCP^r_{s_0,s_1,\ldots,s^r}$, where the weights $s_i$ are 1 and the
coefficients of the highest coroot of $G$ and obey
$\sum_{i=0}^{r}s_i=h$. The supersymmetric condition is also
generalized by using this dual Coxeter number $h$. The corresponding
brane configuration is probably constructed by adding orientifold
planes as like as supersymmetric Yang-Mills theory with orthogonal and
symplectic gauge groups \cite{LLL,BHKL,EGKT}. It would be interesting
to find the correspondence between the index and the more general
$s$-rule including the orientifold planes.

The original computation of the index is for $N{=}1$ Yang-Mills
Chern-Simons theory, which is chiral. It is generally hard to
construct the chiral theory on the branes. Moreover, the Chern-Simons
coefficient of $N{=}1$ supersymmetric theory is renormalized and
shifted by $-\sgn(k)\frac{h}{2}$ \cite{KLL}. This renormalization is closely
related to the derivation of the index. However, we have not
understood the meaning of the renormalization of the coefficient in
terms of the branes in superstring theory. We hope that we can also
analyze the dynamics of lower supersymmetric theory by using branes.

The construction of the dual or mirror theory in terms of the brane
dynamics is simple as mentioned. However, field theoretical meanings
of these symmetries have not been so clear. It is interesting to
extend the analysis of Ref.~\cite{KS} to the non-Abelian case. We hope
that the brane configurations in superstring theory help to understand
non-perturbative dynamics of supersymmetric quantum field theories in
three-dimensions.

\subsection*{Note added}

As I completed this paper, I received Ref.~\cite{BHKK}, which also
considers supersymmetry breaking in $N{=}3$ Yang-Mills Chern-Simons
theory and its $N{=}2,1$ deformations in a similar manner using
branes.

\section*{Acknowledgments}

I would like to thank the organizers of the Summer Institute '99 in
Fuji-Yoshida for hospitality during completion of this work. I am also
grateful to all participants for useful comments and discussions.

\input refs.tex

\end{document}

%% file: refs.tex
%
%

\newcommand{\NP}[3]{Nucl.\ Phys.\ {\bf #1} (#2) #3}
\newcommand{\PL}[3]{Phys.\ Lett.\ {\bf #1} (#2) #3}
\newcommand{\PR}[3]{Phys.\ Rev.\ {\bf #1} (#2) #3}
\newcommand{\PRL}[3]{Phys.\ Rev.\ Lett.\ {\bf #1} (#2) #3}
\newcommand{\JHEP}[3]{JHEP {\bf #1} (#2) #3}
\newcommand{\hepth}[1]{{\tt hep-th/#1}}

\newcommand{\lit}[3]{#1, ``#2,'' #3}